\documentclass[a4paper]{jpconf}
\usepackage{graphicx}
\usepackage[utf8]{inputenc}
\usepackage[T1]{fontenc}
\usepackage{commath}
\begin{document}
\title{Jet--induced modifications of the characteristic of the bulk nuclear matter}

\author{P~Marcinkowski$^1$, M~Słodkowski$^1$, D~Kikoła$^1$, J~Sikorski$^3$, J~Porter-Sobieraj$^2$, P~Gawryszewski$^1$ and B~Zygmunt$^2$}

\address{$^1$ Faculty of Physics, Warsaw University of Technology, Koszykowa 75, 00-662 Warsaw, PL}

\address{$^2$ Faculty of Mathematics and Information Science, Warsaw University of Technology, Koszykowa 75, 00-662 Warsaw, PL}
  
\address{$^3$ Faculty of Physics, University of Warsaw, Pasteura 5, 02-093 Warsaw, PL}

\ead{patryk.marcinkowski@protonmail.ch}

\begin{abstract}
We present our studies on jet--induced modifications of the characteristic of the bulk nuclear matter. To describe such a matter, we use efficient relativistic hydrodynamic simulations in (3+1)dimensions employing the Graphics Processing Unit (GPU) in the parallel programming framework. We use Cartesian coordinates in the calculations to ensure a high spatial resolution that is constant throughout the evolution of the system. We show our results on how jets modify the hydrodynamics fields and discuss the implications.
\end{abstract}


\vspace{-0.7cm}
\section{Simulation setup}
The dynamics of the bulk matter is simulated using the ideal relativistic hydrodynamic equations with a source term  \cite{hirano}
\begin{equation}\label{eqn:hydro}
  \partial_{\mu} T^{\mu\nu}=
  \partial_{\mu}\left((e+p)u^{\mu}u^{\nu}-pg^{\mu\nu} \right)=
  S^{\nu}\\
\end{equation}
where $u=u(x)$ represents a~conserved variable, which include energy, momentum and charge densities. The source term represents energy deposited by jets in the system and it is defined as
\begin{equation}
S^{\nu}(\vec{x})
=\sum_{i=1}^{n_{jet}}\left(-\frac{\mathrm{d} E}{\mathrm{d} t}, -\frac{\mathrm{d} \vec{M}}{\mathrm{d} t}\right)\delta ^{(3)}(\vec{x}-\vec{x_i}(t))
\end{equation}
where $n_{jet}$ is the number of jets.
We implemented several algorithms for solving the problem, including WENO \cite{shu1} ($5^{th}$ and $7^{th}$ order) and Musta-Force \cite{toro}. Currently we use $7^{th}$ order WENO, which has proven high performance and good quality in test problems with steep gradients. The~numerical scheme is integrated over time  using third order Runge-Kutta algorithm.
\\ \indent
The energy loss of jets propagating through the bulk nuclear matter is described by using two mechanisms: radiation of gluons and collisions of partons in dense medium and is given by
\begin{equation}\label{eqn:energy}
\left(-\frac{\mathrm{d} E}{\mathrm{d} x}\right )=
\kappa_{rad}\frac{C_R}{C_F}T^{3}x+\kappa_{coll}\frac{C_R}{C_F}T^{2}
\end{equation}
where $T$ is local temperature and $\kappa_{rad},\kappa_{coll},C_R,C_F$ coefficients depend on jet flavor (quark or gluon) and its energy \cite{solana}. In the simulations, we assumed $\kappa_{rad}=4,\kappa_{coll}=2.5,C_R/C_F=1$. We assume that $\frac{\mathrm{d} E}{\mathrm{d} x}$ is small compared to the jet energy and the jet is not modified by the medium.
\\ \indent
The code is written using NVIDIA CUDA programming framework purposed for parallel computing on graphical processing units (GPU). Our implementation \cite{pub1,pub2,pub3,pub4} speeds up more than 2 orders of magnitude in comparison to the equivalent CPU execution.

\section{Results}
We simulated propagation and energy loss $\left(\frac{\mathrm{d} E}{\mathrm{d} x}\right)$ of a high-energy parton through the medium and analyzed the evolution of the system. The initial condidtions were based on ellipsoidal flow \cite{ellip} with initial simulation time $t_0=1$ fm/c.

\vspace{-0.3cm}
\begin{figure}[h]
\begin{minipage}{16pc}
\includegraphics[width=14pc]{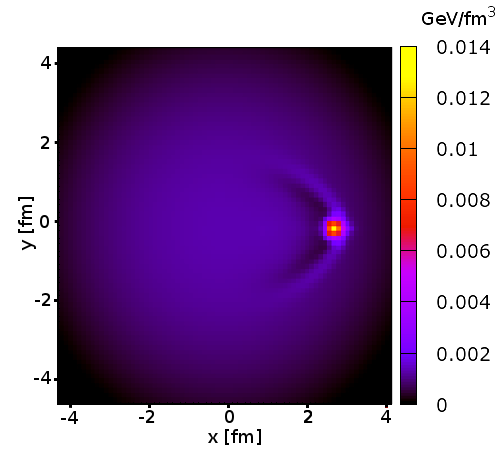}
\caption{\label{energy}Energy density in plane $xy$ at~$z=0$~after $t=2.4$ fm/c.}
\end{minipage}\hspace{4pc}%
\begin{minipage}{14pc}
\includegraphics[width=14pc]{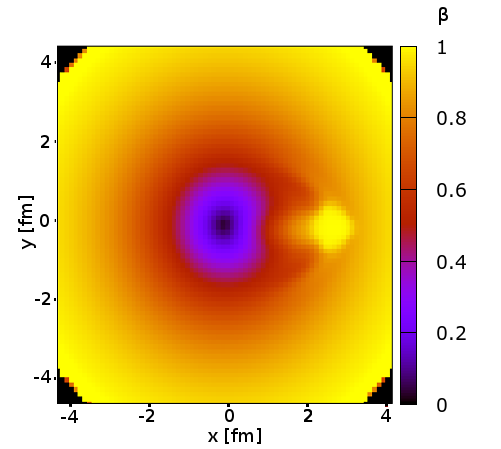}
\caption{\label{velocity}Velocity profile in plane $xy$ at~$z=0$ after $t=2.4$ fm/c.}
\end{minipage} 
\end{figure}

\vspace{-0.8cm}
\begin{figure}[h]
\includegraphics[width=16pc]{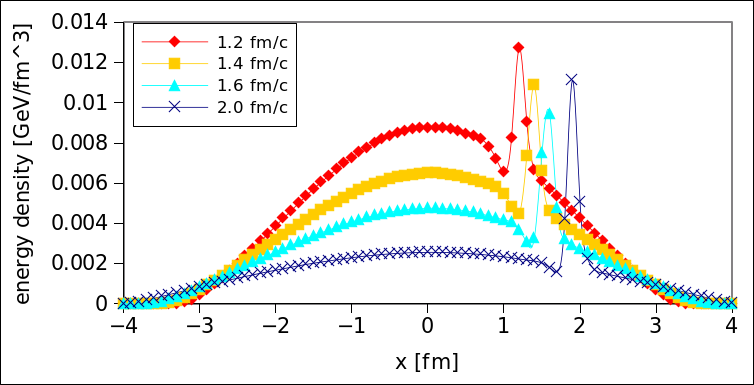}\hspace{2pc}%
\begin{minipage}[b]{16pc}
\hspace{4pc}%
\caption{\label{evo}Evolution of the~system: energy density 1D sections at $y=z=0$ at different points of time (from 1.2 fm/c to 2 fm/c after simulation start).\\
Simulation parameters: 
d$x=0.1$ fm, 
d$t=0.02$ fm/c, 
grid size $256\times 256\times 256$,
EOS $p=e/3$.}
\end{minipage}
\end{figure}

\vspace{-0.3cm}
In figures \ref{energy} and \ref{velocity}, there is a clean signal of jet propagation in the medium: a Mach cone is visible both in energy and velocity distributions. Figure \ref{evo} shows the evolution of energy density with time. There is a clear indication that jet energy loss has a significant impact on the~system. The area where the jet interacted with system has much larger energy density compared to the~surrounding matter and the difference increases with time.
\\ \indent
In the future we plan to employ more realistic jet energy loss algorithm and add freeze--out to study the effect on the experimental observables: elliptic flow and higher flow harmonics. 

\section*{References}
\begin{thebibliography}{11}

\bibitem{hirano} Tachibana Y and Hirano T 2014 {\it Phys. Rev.} C {\bf 90} 021902

\bibitem{shu1} Zhang X and Shu C W 2012 {\it J. Comput. Phys.} {\bf 231}(5) 2245-58

\bibitem{toro} Toro E F 2006 {\it Appl. Numer. Math.} {\bf 56}(10-11) 1464-79

\bibitem{solana} Casalderrey-Solana J, Gulhan D C, Milhano J G, Pablos D and Rajagopal K  2014 {\it J. High Energy Phys.} JHEP10(2014)019

\bibitem{pub1} Porter-Sobieraj J, Cygert S, Kikoła D, Sikorski J and Słodkowski M 2015 {\it Concurrency and Computation: Practice and Experience} {\bf 27}(6) 1591-602

\bibitem{pub2} Cygert S, Kikoła D, Porter-Sobieraj J, Sikorski J and Słodkowski M 2014 {\it Lect. Notes Comput. Sc.} {\bf 8384} 500-9

\bibitem{pub3} Sikorski J, Cygert S, Porter-Sobieraj J, Słodkowski M, Krzyżanowski P, Książek N and Duda P 2014 {\it. J. Phys. Conf. Ser.} {\bf 509} 012059

\bibitem{pub4} Cygert S, Porter-Sobieraj J, Kikoła D, Sikorski J and Słodkowski M 2013 {\it Fed. Conf. on Computer Science and Information Systems (Kraków)} pp 441-6

\bibitem{ellip} Sinyukov Y M and Karpenko I A 2005 {\it Preprint} nucl-th/0505041

\end{thebibliography}
\end{document}